\documentclass[useAMS,usenatbib,usegraphicx]{mn2e}

\usepackage{psfig}
\usepackage{times}
\usepackage{subfigure}

\newcommand{\kms}{\hbox{km s$^{-1}$}}
\newcommand{\ms}{\hbox{m s$^{-1}$}}

\newcommand{\vsini}{\hbox{$v$\,sin\,$i$}}

\newcommand{\degs}{$\degr$}
\newcommand{\chisq}{$\chi^{2}$}

\newcommand{\radday}{\hbox{rad.day$^{-1}$}}
\newcommand{\invday}{\hbox{day$^{-1}$}}

\newcommand{\ha}{H$\alpha$}

\newcommand{\hkaqr}{\hbox{HK Aqr}}

\newcommand{\lopeg}{\hbox{LO Peg}}

\begin{document}

\title[LO Peg in 1998]{LO Peg in 1998: Starspot patterns and differential rotation}

\makeatletter
 
\def\newauthor{%
  \end{author@tabular}\par
  \begin{author@tabular}[t]{@{}l@{}}}
\makeatother

\author[J.R.~Barnes, A.~Collier~Cameron, T.A.~Lister, G.R.~Pointer, M.D.~Still]
{J.R.~Barnes,$^1$\thanks{E-mail: jrb3@st-andrews.ac.uk} A.~Collier~Cameron,$^1$ T.A.~Lister,$^1$ G.R.~Pointer,$^1$ M.D.~Still$^2$ \\
$^1$ School of Physics and Astronomy, University of St Andrews, Fife KY16 9SS. UK. \\
$^2$  Universities Space Research Association, NASA/Goddard Space Flight Center,  Code 662, Greenbelt, MD 20771, USA}

\date{2004, 200?}

\maketitle

\begin{abstract}

We present Doppler images of the young K5V - K7V rapid rotator LO Peg from seven nights of continuous spectroscopy obtained in 1998 from July 04 to July 10. The images reveal the presence of a strong polar cap with appendages extending to mid-latitudes, but no starspots are seen below 15\degr. We briefly discuss the distribution of spots in light of recent flux transport simulations which are able to reproduce the observed latitude dependence.

With the full timeseries of spectra, of which 314 are useful, many phases are observed three times over the seven nights of observations.  Using starspots as tracers of a solar-like latitudinal differential rotation in our image reconstructions, we find that the equatorial regions complete one more rotation than the polar regions every \mbox{181 $\pm$ 35} d. LO Peg is the second coolest star for which such a measurement has been made using indirect imaging methods. The degree of latitudinal shear is less than that seen in G and early K dwarfs, suggesting a trend in which differential rotation decreases with stellar mass in (pre-)main sequence objects.

\end{abstract}


\section{INTRODUCTION}

Several methods which enable us to measure the surface differential rotation in late-type stars with convective envelopes have appeared in the literature. Variability of stellar activity diagnostics such as the Ca {\sc ii} H \& K emission cores has been used to determine stellar rotation periods as exemplified by the long term Mount Wilson Survey \citep{wilson78,baliunas95}. In many stars, the measured rotation period is found to vary periodically from one epoch to another over timescales similar to the 11 year solar magnetic activity cycle. If we adopt the solar paradigm, in which changes in latitude of the active regions occur throughout the cycle, we can attribute the measured period changes to a latitude dependent rotation in stars. 
The scatter in measured periods gives some indication of the degree of differential rotation shear on the star \citep*{donahue96}.
 
A similar method based on photometric modulation due to starspots, in a sample of stars, has been found to yield a trend in the degree of surface shear with rotation rate \citep{hall91dynamo,henry95diffrot}. Long term variations of the photometric period been shown to be indicative of magnetic activity cycles \citep{messina02}. More recently, \citet{reiners03diffrot} have found that main-sequence stars of spectral types F and G  exhibit high degrees of differential rotation. The rotational shear manifests itself in the Fourier transforms of the rotationally broadened line profiles. This allows precise measurements of shear from line profiles alone for stars with intermediate rotation rates.

A further method of determining the latitude dependence of rotation is to use spot distributions from indirect imaging techniques. Here we do not have to make assumptions about the latitude extent of starspots in order to determine the magnitude of shear between the equator and pole. We are however restricted to observing rapid rotators if we are to obtain sufficient resolution elements across a rotationally broadened profile to make this technique viable. The number of resolution elements is determined by a number of stellar broadening mechanisms, but is ultimately limited by the instrumental resolution. By cross-correlating constant latitude strips in Doppler images derived at closely related epochs, the latitude dependent surface shear can be determined \citep{donati97doppler}. Alternatively, a differential rotation law (usually assumed to be solar-like) can be incorporated into the imaging model, and the whole data set from several nights combined to derive an image (\citealt{donati00rxj1508}; \citealt*{petit02}).

\begin{table*}
\begin{minipage}{135mm}
\caption{Journal of observations of LO Peg and standards on 1998 July 04 - 10}
\protect\label{tab:obs_journal}
\vspace{5mm}
\begin{center}
\begin{tabular}{lccccl}
\hline
Object          & UT Start    	& UT End      		& Exp time [s]   	& No. of frames& Comments		\\
\hline

		& 		&               		&  1998 Jul 04 &              	&			\\
\vspace{-3mm} \\
Gl 673		& 22:27	& 			& 100		& 3		& K5V template          \\
LO Peg	& 01:32	& 05:57		& 200		& 49		&			\\
\hline
                	& 		&              	 	& 1998 Jul 05	&              	&			\\
\vspace{-3mm} \\
Moon		& 21:57	& 22:07		& 200, 30, 60   & 3		& Solar template        	\\
LO Peg	& 01:46	& 05:56		& 200		& 47		&			\\
\hline
                	& 		&               		&  1998 Jul 06 &              	&			\\
\vspace{-3mm} \\
Gl 687		& 21:38	& 21:53	        	& 300, 200, 200& 3		& M3V template	\\
Gl 649		& 22:03	& 22:17		& 200		& 3		& M1V template	\\
LO Peg	& 01:40	& 05:54		& 200		& 47		&			\\
\hline
                	& 		&               		& 1998 Jul 07  &              	&			\\
\vspace{-3mm} \\
LO Peg	& 01:46	& 06:01		& 200		& 48		&			\\
\hline
                	& 		&               		& 1998 Jul 08	&              	&			\\
\vspace{-3mm} \\
LO Peg	& 01:34	& 05:59		& 200		& 50		&			\\
\hline
                	& 		&               		& 1998 Jul 09  &              	&			\\
\vspace{-3mm} \\
LO Peg	& 01:51	& 06:00		& 200		& 47		&			\\
\hline
                	& 		&               		& 1998 Jul 10  &              	&			\\
\vspace{-3mm} \\
Gl 447		& 20:49	& 21:23		& 600		& 3		& M4V template 	\\
LO Peg	& 01:41	& 06:00		& 200		& 47		&			\\
\hline 

\end{tabular}
 \end{center}
\end{minipage}
\end{table*}

LO Peg (BD+22 4409) is among the least massive of the young rapid rotators in the solar neighbourhood. It was detected by the {\em ROSAT WFC EUV} all-sky survey as the source RE J2131+23, and by the {\em Extreme Ultraviolet Explorer} survey as the source EUVE J2131+23.3 \citep{malina94}. \citet{jeffries93kinematics} identified LO Peg as a member of the Local Association on the basis of its Galactic space motions and a large EUV to bolometric flux ratio. LO Peg was first studied in detail by \citet{jeffries94bd22d4409} who determined an axial rotation period of \hbox{0.42375 d} from V-band photometric observations. It was also noted that a slightly shorter period of \hbox{0.38417 d} gave a physically acceptable solution with low false alarm probability and folded lightcurve with little scatter. Photometry yielded a visual magnitude of \mbox{$V = 9.19 \pm 0.05$} and colours \mbox{$(B-V) = 1.08 \pm 0.02$}, \mbox{$(U-B) = 0.82 \pm 0.08$}, \mbox{$(V-R)_{KC} = 0.59 \pm 0.02$} and \mbox{$(V-I)_{KC} = 1.23 \pm 0.03$}. Study of the Lithium 6708 \AA~line yielded an abundance of N(Li) = {$1.30 \pm 0.25$}, while the iron abundance [Fe/H] = \mbox{$0.0 \pm 0.1$}. The equatorial rotation velocity was determined to be \vsini~= \mbox{$69 \pm 1$} \kms. \citet{jeffries94bd22d4409} also obtained a model fit to the spectrum of LO Peg, and found a spectral type of K5 - K7 with an age in excess of 30 Myr.

\begin{figure*}
\begin{center}

  \begin{tabular}{cc}
    \includegraphics[width=5.30cm,angle=0]{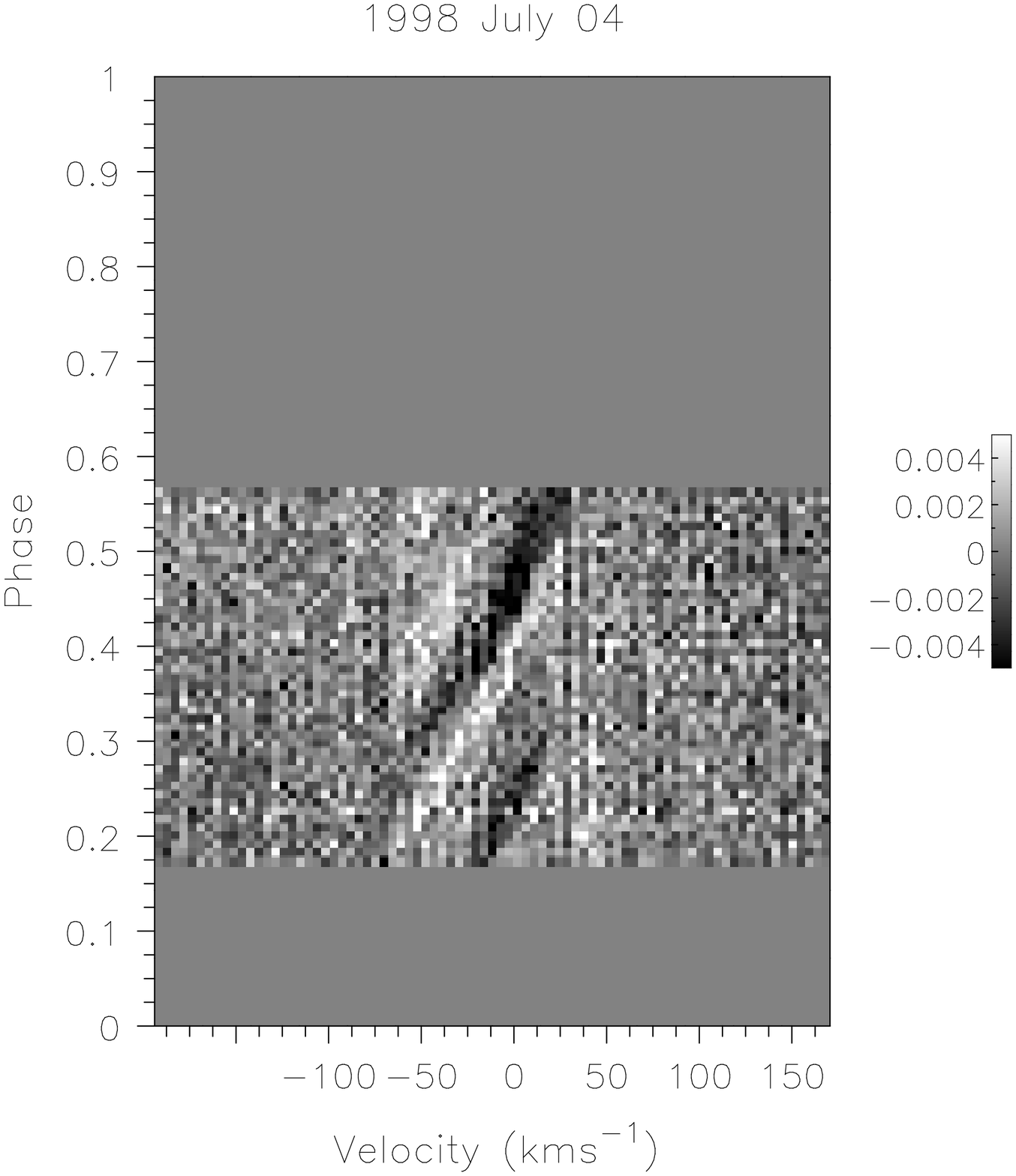} &
    \includegraphics[width=5.30cm,angle=0]{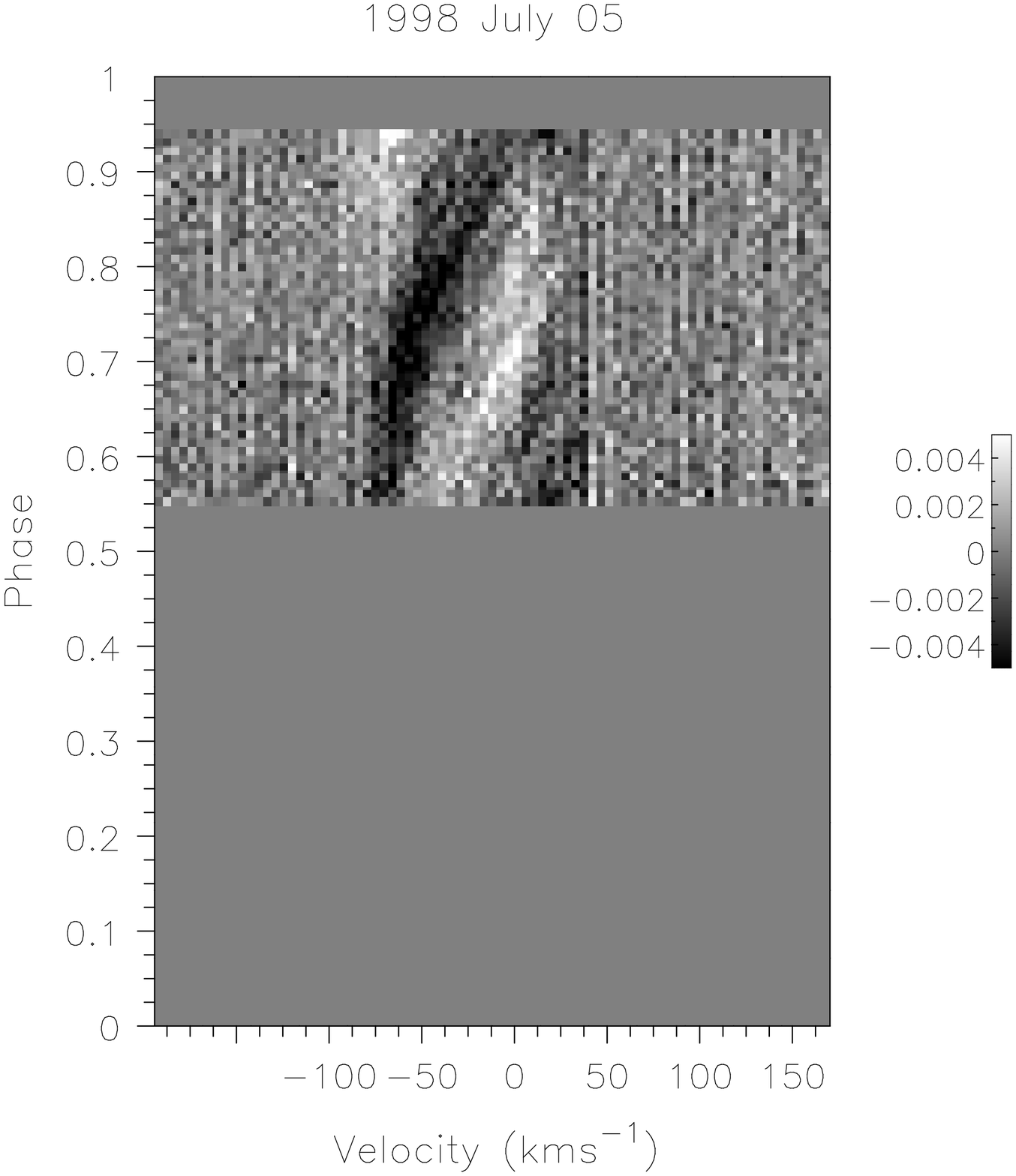} \\
  \end{tabular}
\smallskip
  \begin{tabular}{cc}
    \includegraphics[width=5.30cm,angle=0]{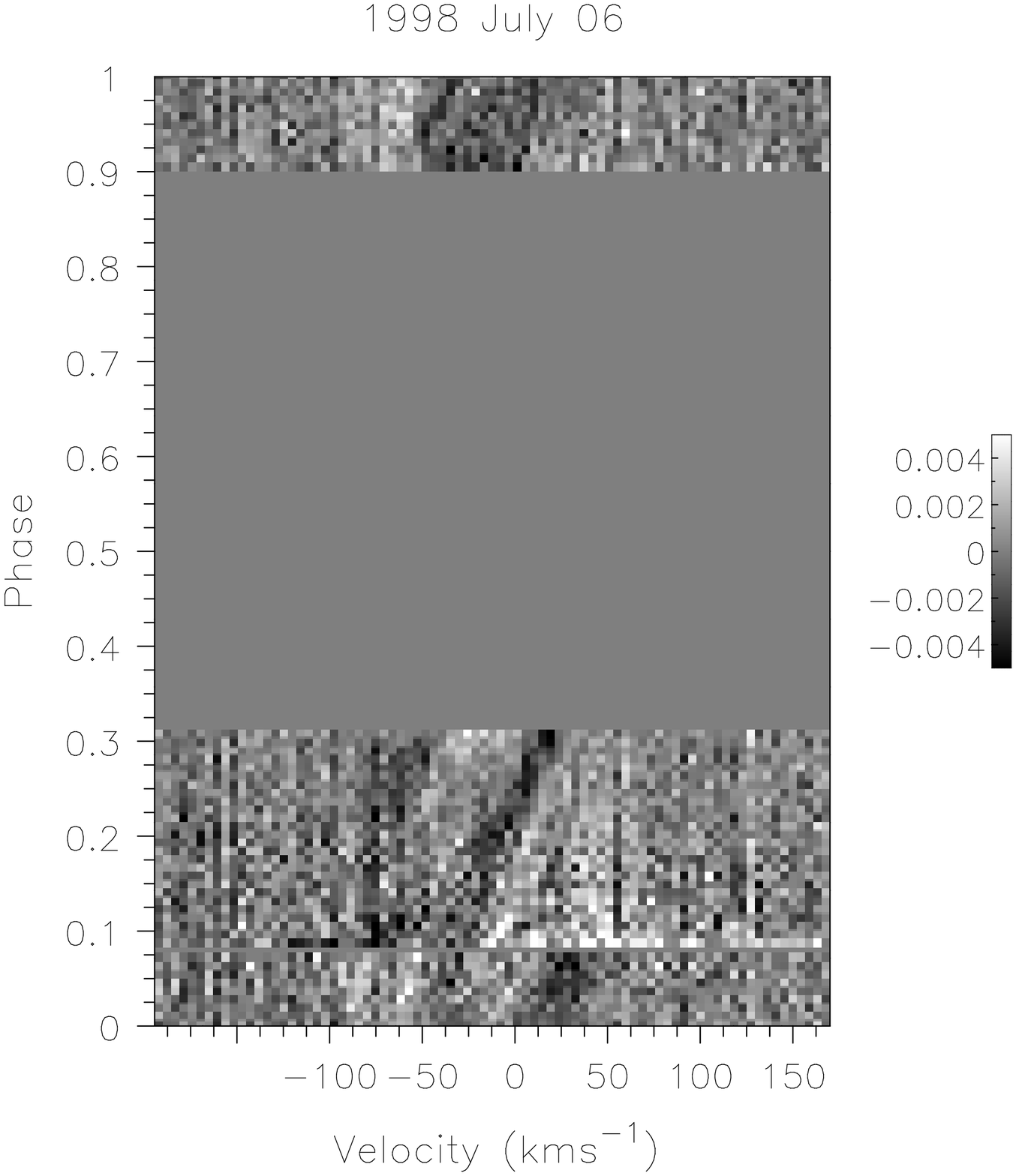} &
    \includegraphics[width=5.30cm,angle=0]{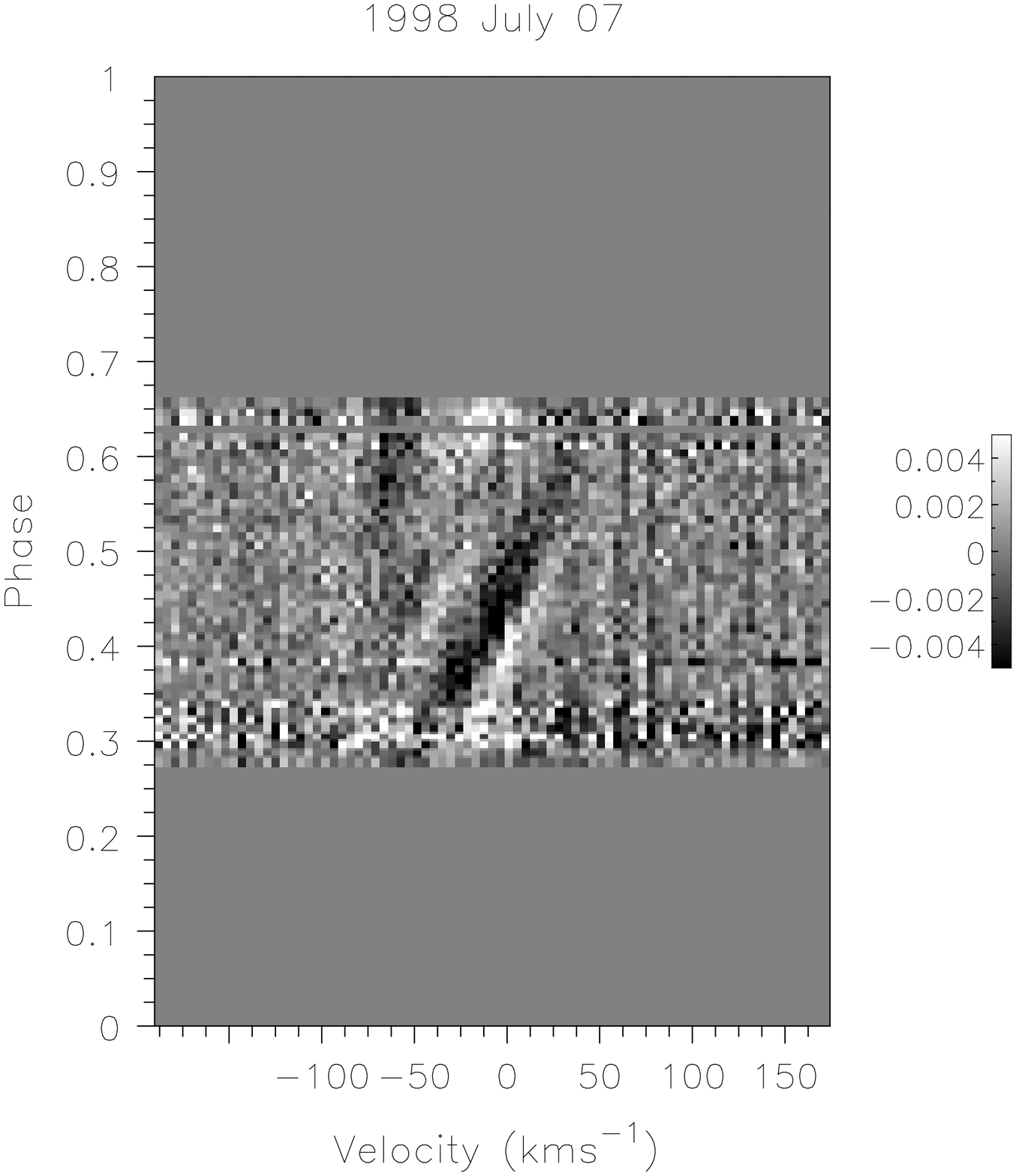} \\
  \end{tabular}
 \smallskip
  \begin{tabular}{ccc}
    \includegraphics[width=5.30cm,angle=0]{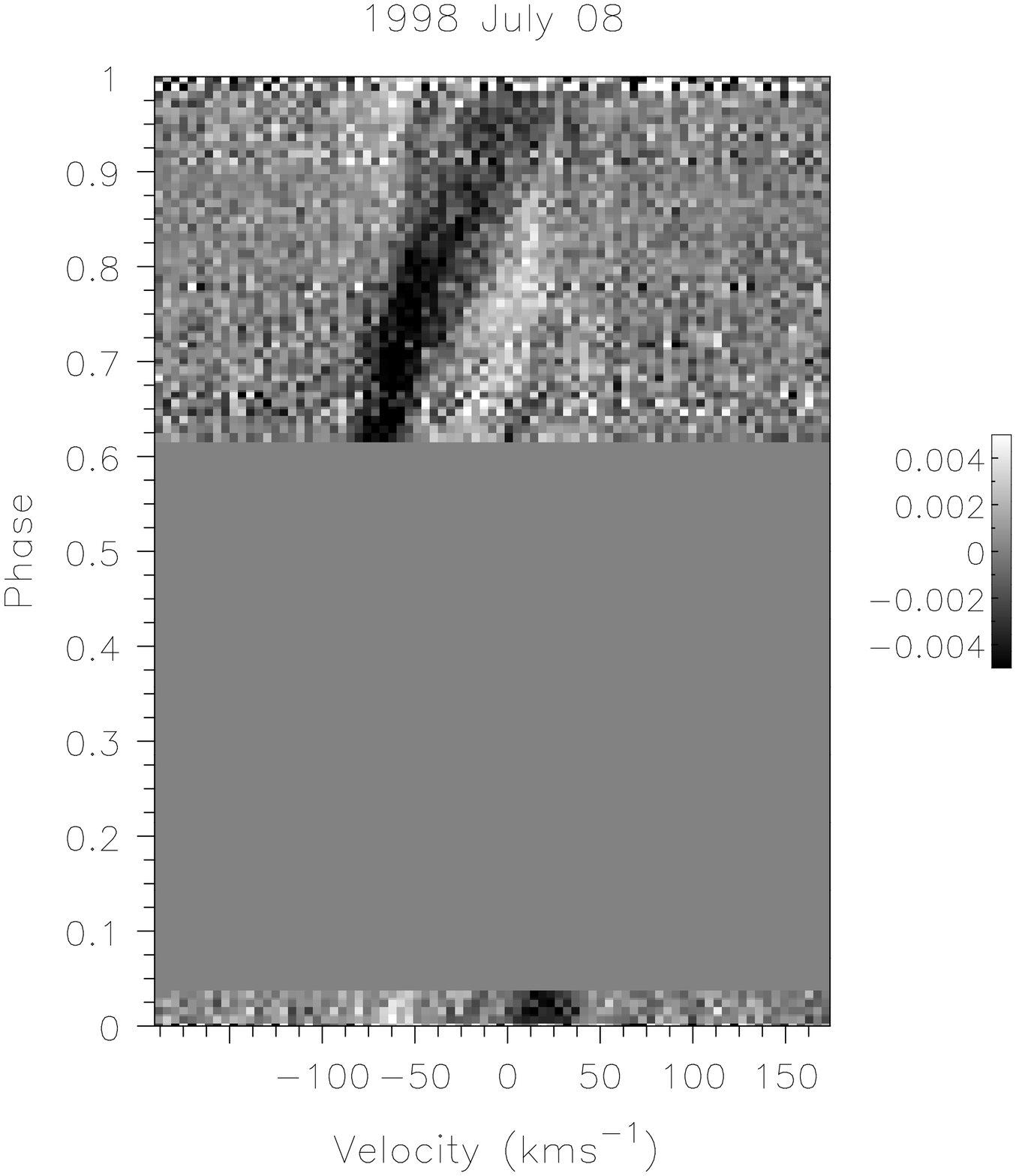} &
    \includegraphics[width=5.30cm,angle=0]{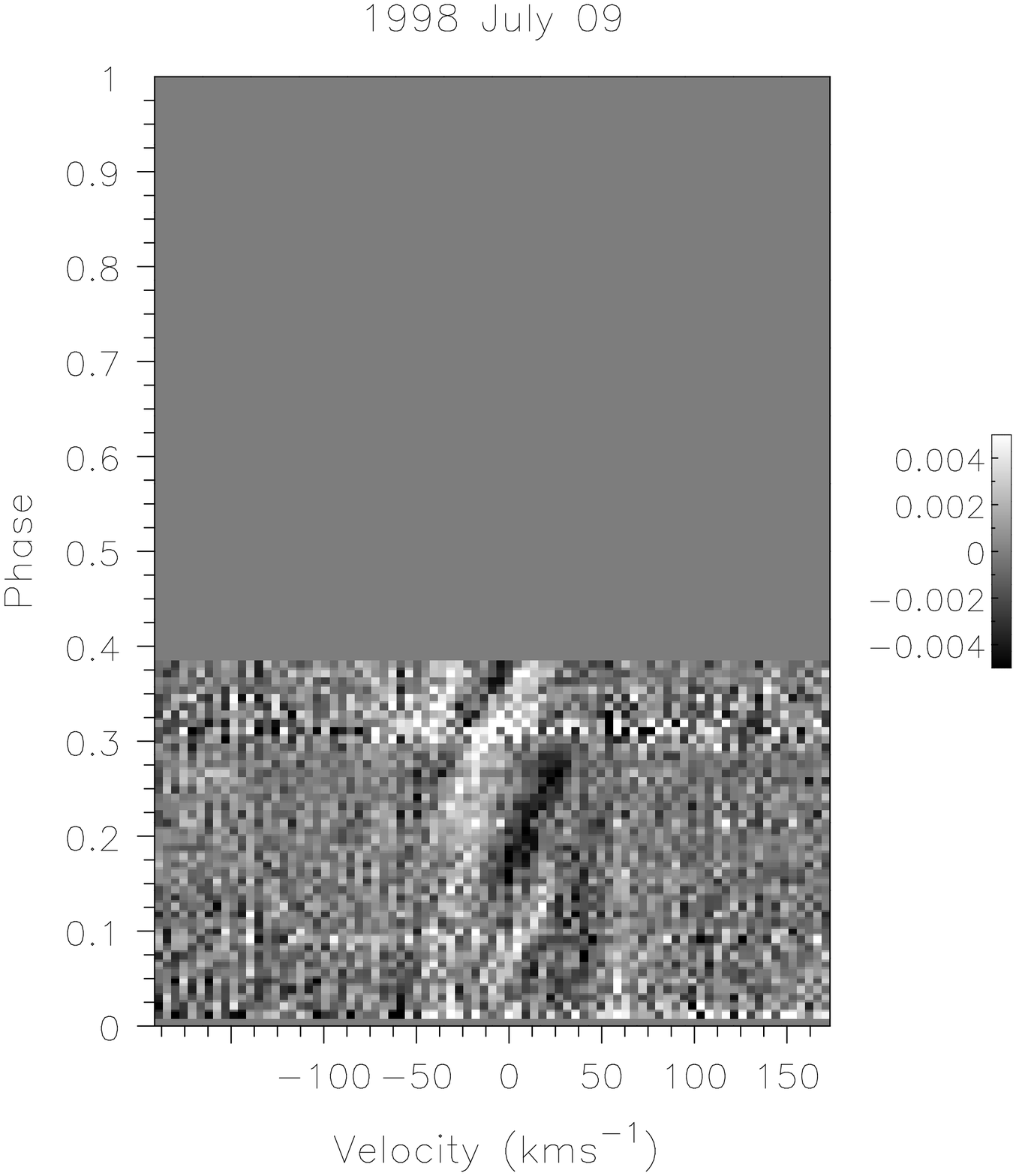} &
    \includegraphics[width=5.30cm,angle=0]{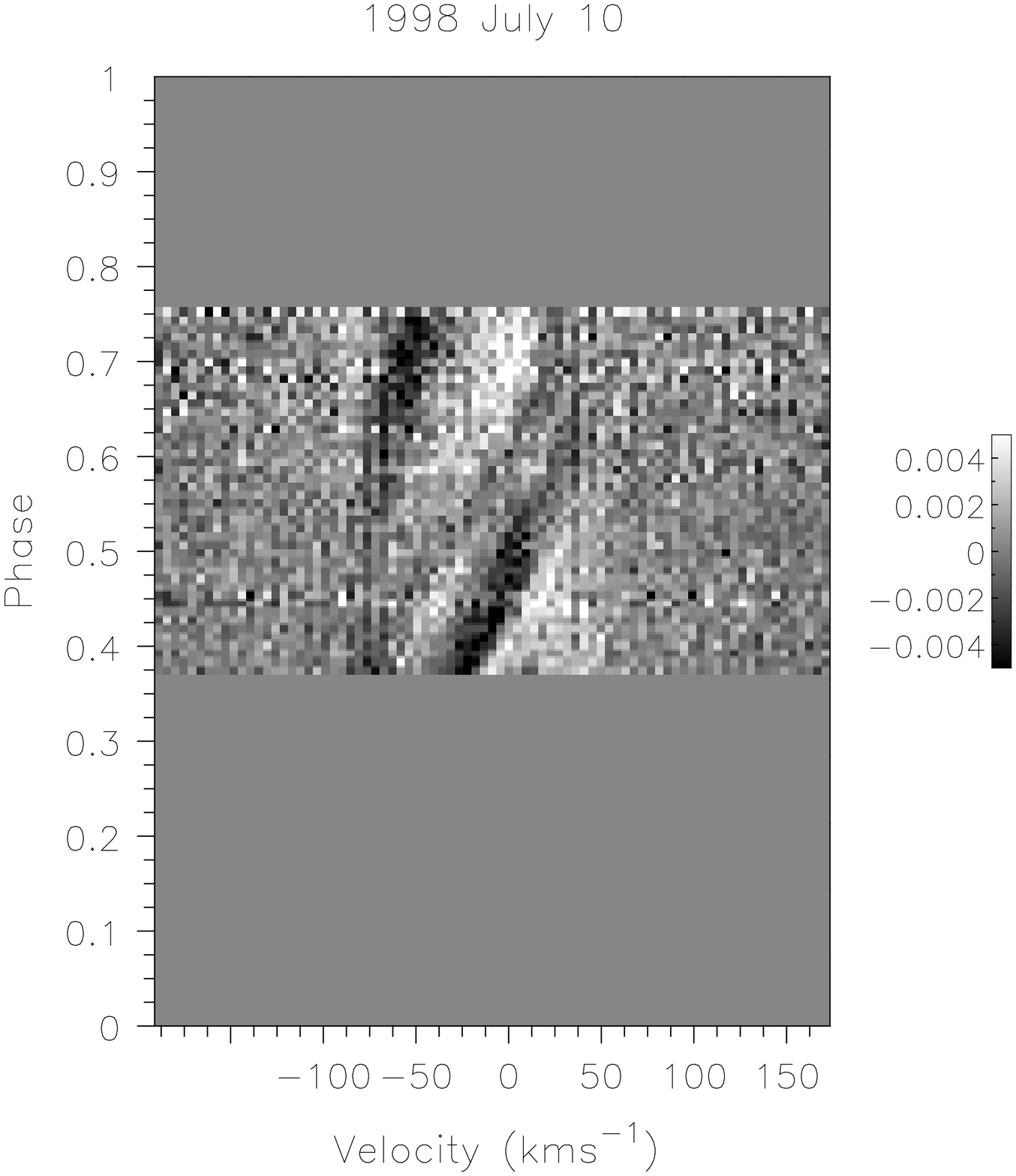} \\
  \end{tabular}

\end{center}

\caption[LO Peg time series spectra ]{Time series spectra for 1998 July 04\,-\,10, with phase plotted against velocity. The mean profile has been subtracted from each time series. White features correspond to starspot signatures. }
\label{lopeg_spectra}
\end{figure*}

\citet{jeffries94bd22d4409} also determined the axial inclination of LO Peg to be 50\degs. By studying the \ha\ line, which appears in emission, these authors showed that LO Peg does not exhibit prominence features, which appear as \ha\ absorption transients in other similar stars as the cool optically thick clouds cross the stellar disc \citep{cameron89cloud}. The lack of observational evidence indicates that if prominences do exist, they must be present only above low latitudes where they will not cross the stellar disc in the observer's line of sight. More recently, \citet{eibe99bd} found that variability of the \ha\ line indicates strong downflows of absorbing material. A previous imaging study by \citet*{lister99lopeg} has yielded surface Doppler images showing a polar cap and a low latitude band of features on the surface. Observations at two epochs allowed these authors to distinguish between the two proposed axial rotation periods discussed above, indicating that the longer 0.42375~d period is preferred.

The K5V\,-K7V spectral type make LO Peg an important object because no other single stars of this mass have been studied. In fact, no stars between K3V and  and M1V have been imaged to date. At this spectral type, the radiative core decreases in size as the star approaches the fully convective regime at spectral type early to mid-M. If a solar-like interface dynamo is at work, according to \citet{schussler96buoy} and \citet{granzer00} we may expect only intermediate to high { latitude eruption of magnetic flux}. In this paper, we present images derived from data obtained over seven nights and determine the differential rotation for LO Peg.

\section{OBSERVATIONS AND DATA REDUCTION}

Observations were made with the Multi SIte Continuous Spectroscopy (MUSICOS) instrument { (resolution $\sim$ 30000)}  \citep{baudrand92musicos} at the Isaac Newton Telescope on 1998 July 04 - 10 (Table \ref{tab:obs_journal}). The seeing conditions were generally good at $\sim 1$\arcsec\ with attenuation due to atmospheric dust on the nights of July 07 \& 08. The 2K$\times$2K SITe1 CCD was used in conjunction with the Cassegrain fibre fed spectrograph, giving a useful wavelength coverage of 4359 \AA\ to 7352 \AA\ in which a total of 51 orders were recorded. 

Pixel to pixel variations were removed using flat-field exposures taken with an internal tungsten reference lamp. The worst cosmic ray events were removed at the pre-extraction stage using the FIGARO routine {\sc bclean} \citep{shortridge93figaro}. Scattered light was modelled by fitting polynomials of degree 7 to the sets of inter-order pixels at each X-position in each frame. The spectra were extracted using ECHOMOP, the \'{e}chelle reduction package developed by \citet{mills92}. The Thorium-Argon arc-frames used for wavelength calibration were extracted in conjunction with a target spectrum, and calibrated using this package. The orders were extracted using ECHOMOP's implementation of the { optimal} extraction algorithm developed by \citet{horne86extopt}. ECHOMOP propagates error information based on photon statistics and readout noise throughout the extraction process. \\

\begin{figure*}
\subfigure[]{\includegraphics[height=152mm,angle=270,bbllx=70,bblly=0,bburx=355,bbury=800]{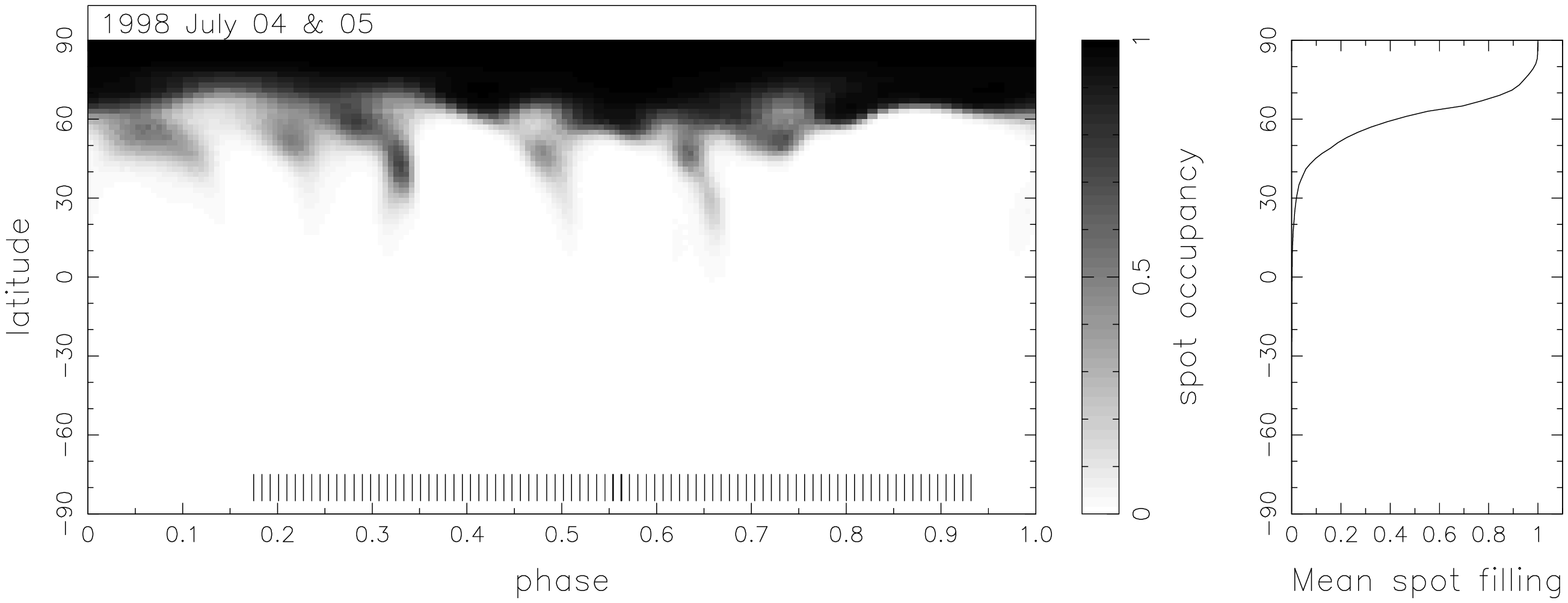} }
\subfigure[]{\includegraphics[height=152mm,angle=270,bbllx=70,bblly=0,bburx=355,bbury=800]{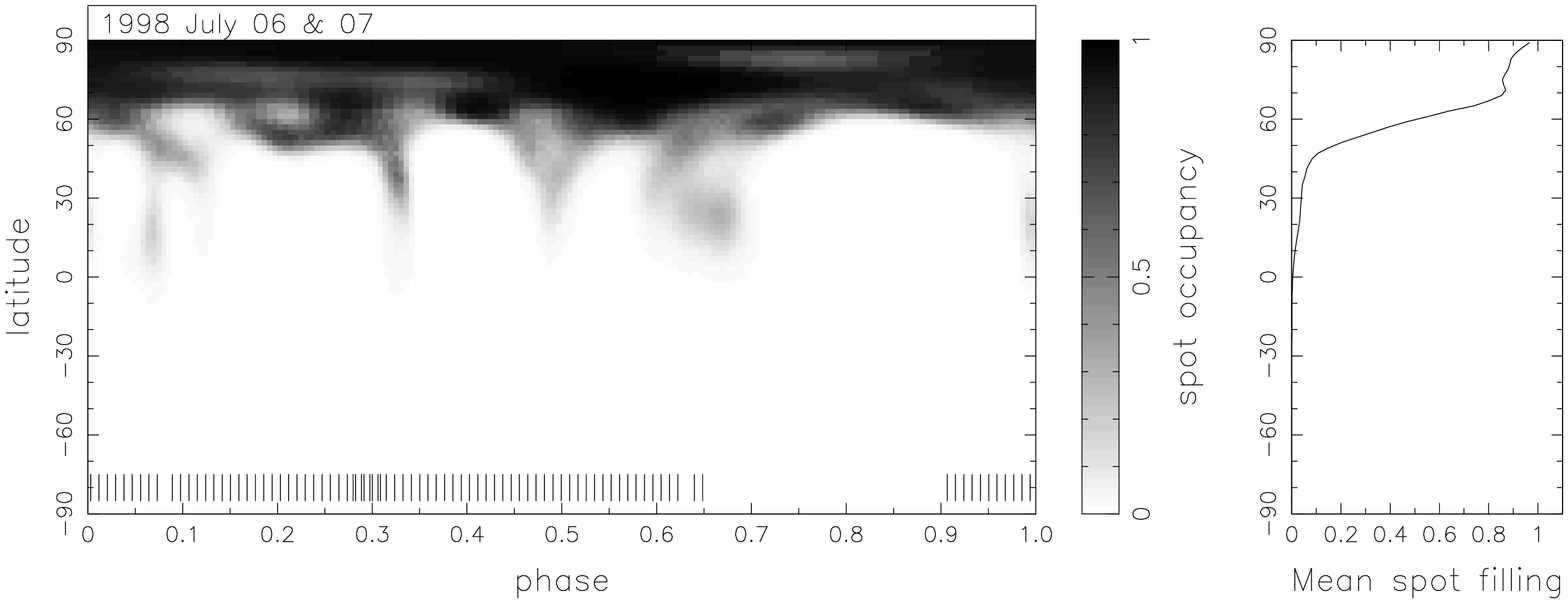} }
\subfigure[]{\includegraphics[height=152mm,angle=270,bbllx=70,bblly=0,bburx=355,bbury=800]{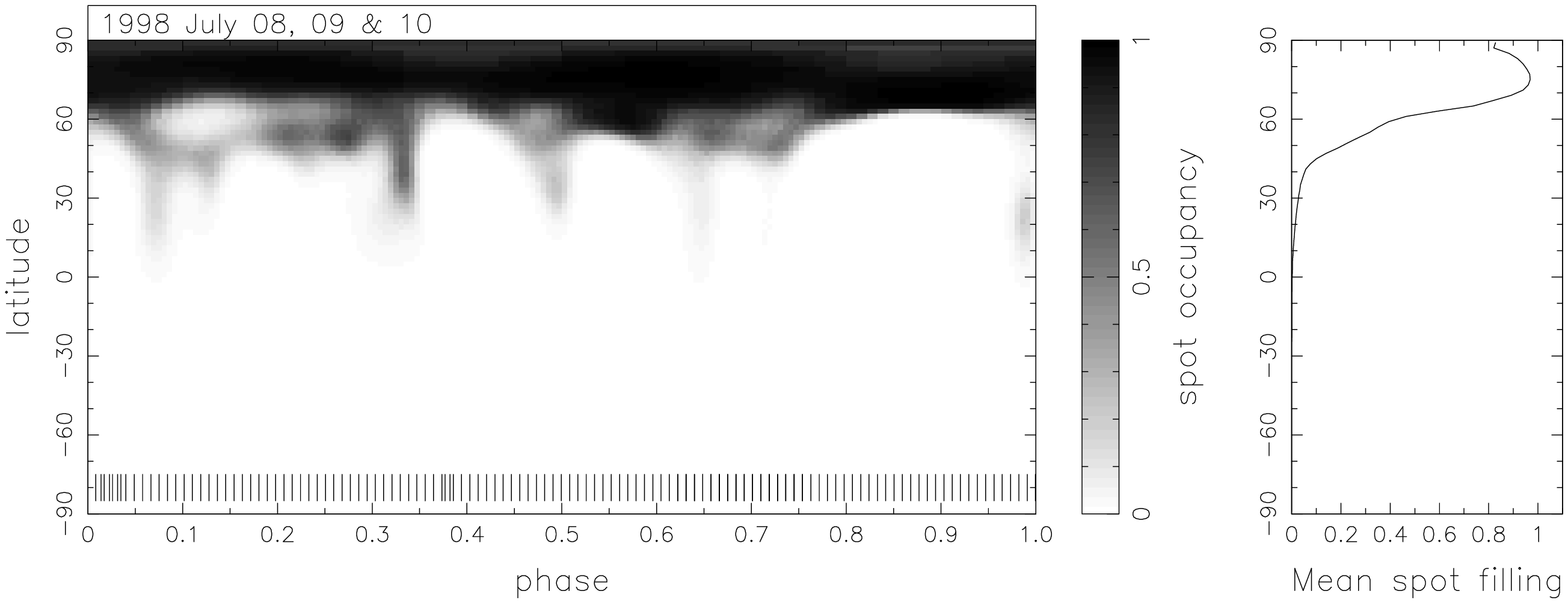} }
\caption[\hkaqr\ reconstructed images]{Maximum entropy regularised reconstruction of \lopeg\ for (a) 1998 July 04 \& 05, (b) July 06 \& 07 and (c) July 08, 09 \& 10.}
\protect\label{fig:lopeg_images}
\end{figure*}

\begin{figure*}
\subfigure[]{\includegraphics[height=152mm,angle=270,bbllx=70,bblly=0,bburx=355,bbury=800]{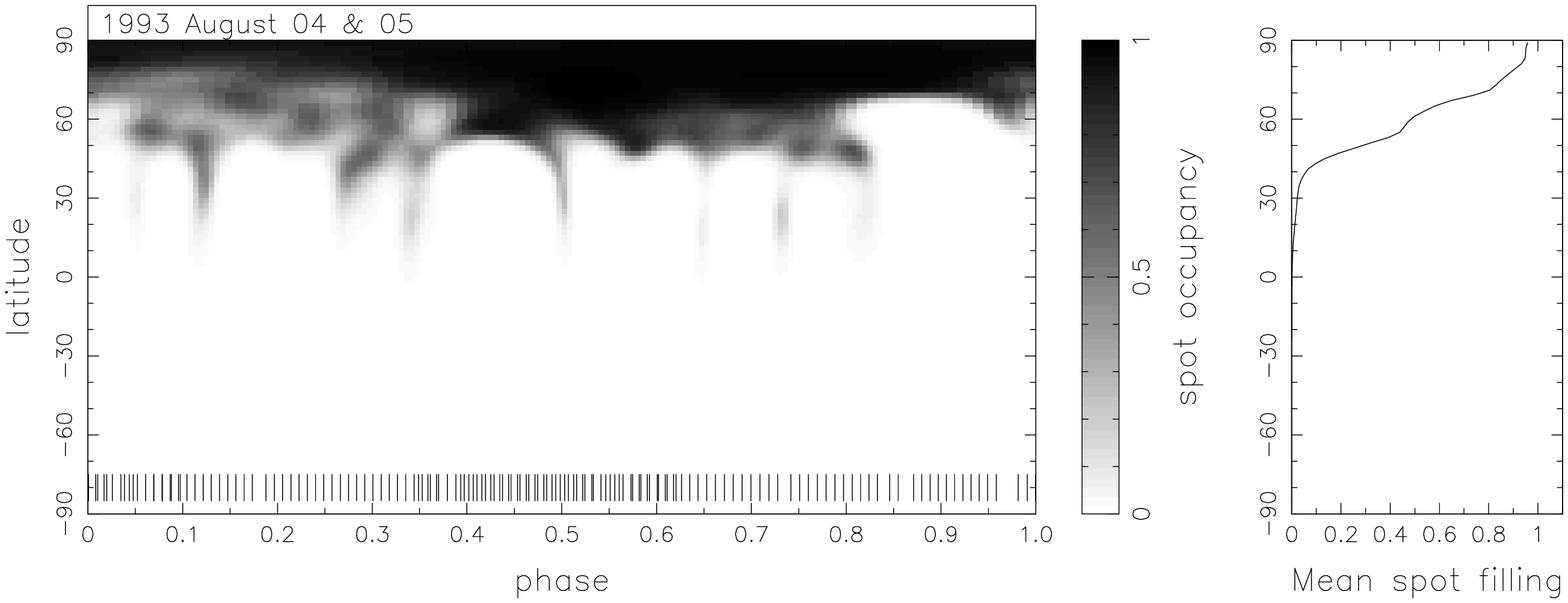}}
\subfigure[]{\includegraphics[height=152mm,angle=270,bbllx=70,bblly=0,bburx=355,bbury=800]{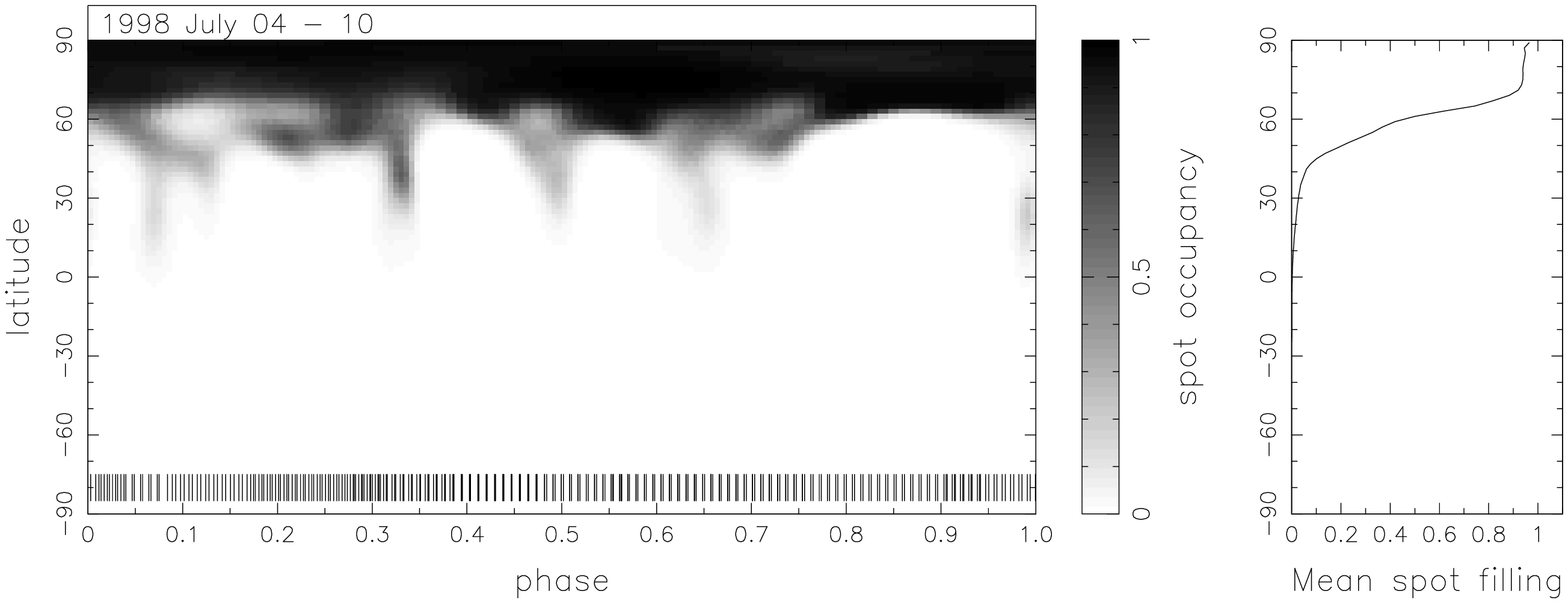}}
\subfigure[]{\includegraphics[height=152mm,angle=270,bbllx=70,bblly=0,bburx=355,bbury=800]{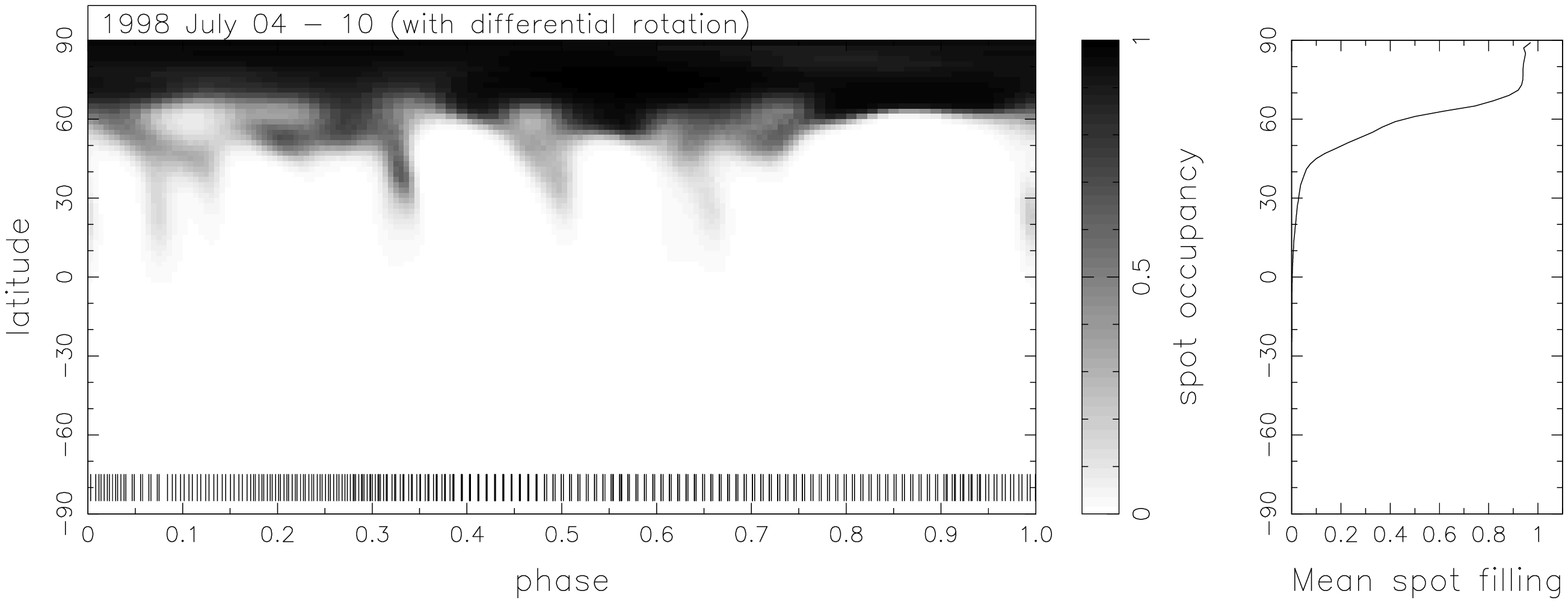}}
\caption[\hkaqr\ reconstructed images]{Maximum entropy regularised reconstruction of \lopeg\ for
(a) 1993 August 04 \& 05, (b) 1998 all nights combined without differential rotation and (c) 1998 all nights combined {\em with} differential rotation.}
\protect\label{fig:lopeg_images2}
\end{figure*}

\subsection{Least squares deconvolution}

We apply least squares deconvolution \citep{donati97zdi} in order to obtain sufficient S/N in the rotation profile at each phase, enabling us to reliably and consistently reconstruct surface images. Due to extreme vignetting in the faintest (blue) orders, we applied a mask to exclude the regions where few counts were recorded. Nevertheless, 3897 images of 2513 lines (due to overlapping wavelength ranges in adjacent orders) were used to derive single high S/N ratio profiles. A mean input S/N ratio of 15 was boosted by a multiplex gain of 50, yielding mean absorption profiles with S/N of typically 750 for each observation. { We used a velocity bin size of 4.5 \kms\ chosen to match the CCD pixel size at the mean weighted central wavelength of the deconvolved line (5399 \AA).}

\subsection{System parameters}

Stellar system parameters were determined { empirically from the full data set by searching for the combination of parameters which minimised \chisq (Table \ref{tab:sysparam}). We discuss the results further in Section \ref{section:comp} in light of differences with the values found for these parameters by other authors.}

\section{RESULTS}

\subsection{The Images}

The maps of LO Peg from 1998 July were produced with our image code, DoTS (Doppler Tomography of Stars) \citep{cameron01mapping} and are presented in Figure \ref{fig:lopeg_images}. Since the phase coverage in a single night was approximately 0.4, we have combined either two or three nights in order to obtain images with good phase coverage. This allows us to compare image consistency and to check for surface evolution of starspots. Immediately obvious in all the images is the very strong polar cap which shows reproducible morphology from one night to the next. Since the axial inclination of LO Peg is 45\degs, stellar latitudes above this value are visible at all times. In essence this ensures reasonable reliability of reconstructed starspots above 45\degs\ for phases where no observations were made, although projection effects and limb darkening may mean that the morphology is less well constrained than for well observed phases. Although the polar cap reaches a radius of 25\degs\ at some longitudes, it is by no means uniform. In this respect, there is evidence for lower spot filling, very close to the pole in the July 06 \& 07 and July 08, 09 \& 10  images. The high latitude spot filling may possibly be due to two large spots or spot groups located approximately 180\degr\ apart, and centred around \hbox{phases 0.0 to 0.1} \hbox{and 0.5 to 0.6.}

Below latitude 60\degs, a number of features are repeated in all the images. Intriguingly, most of the starspot structure at this time appears at, or above 45\degs\, with only a few small spots or spot groups below this latitude. In fact, no spots are seen below 15\degs\ at the time of observations. The mean spot latitudinal filling plot on the right of each image shows no real evidence for preferred latitudes below the obvious polar feature as is often seen on other stars of earlier spectral type (e.g. \citealt{barnes01aper}).

\subsection{Comparison with 1993 images of LO Peg}
\protect\label{section:comp}

\begin{table}

\caption[System parameters]{System parameters for LO Peg.}
\protect\label{tab:sysparam}
\vspace{5mm}
\centering
\begin{tabular}{lc}
\hline                     
P [d]				&  0.423229 $\pm$ 0.000048   \\
$v_{r}$~[kms$^{-1}$]		&  -20.90 $\pm$ 0.25   \\
\vsini 	~[kms$^{-1}$]	  	&  65.84 $\pm$ 0.06 \\
Axial inclination [deg]		&  45.0 $\pm$ 2.5   \\
\hline
\end{tabular}
\end{table}

Our images show similarities with those presented by \citet{lister99lopeg} while at the same time revealing marked differences. The polar cap is somewhat weaker, but shows the same non-uniformity as in the 1993 images. The most notable difference is in the location of the low latitude features in 1993 which are confined to a definite band centred \hbox{around 20\degs.}

Caution is needed however when comparing images from the two epochs. \citet{lister99lopeg} derived a significantly different \vsini\ for LO Peg, of 69 \kms\, in agreement with the measurement made by \citet{jeffries94bd22d4409}.   Lister et al. initially estimated \vsini\ by minimising \chisq\ with respect to \vsini\ for an unspotted stellar image. They also performed reconstructions (to a fixed target \chisq) at a range of \vsini\ values, and selected the value that minimised the total area of spots in the image. { Use of initial \chisq~can only be expected to give an approximate fit to the data because an unspotted star yields model profiles which neglect the effects of starspots. Starspots can alter the shape of the profile, leading to incorrect determination of parameters such as radial velocity, equivalent width and \vsini, if their effects are not considered. Radial velocity is the least sensitive parameter in this respect since the mean profile of a timeseries covering a complete rotation cycle will be symmetric. By contrast, when a strong polar spot is present, as is found on LO Peg, the rotationally broadened profile becomes either flat bottomed or more rounded, with a reduced equivalent width. By fitting the profiles derived from an unspotted stellar image, we could expect to find an optimum fit where the line equivalent width is too low and the \vsini\ too high.

Fitting the model to the data for a fixed level of \chisq\, and measuring the starspot area (a proxy for the image entropy), may also lead to incorrect choice of system parameters. If a fixed \chisq\ is chosen, the combinations of parameters yielding the minimum spot area may occur before many distortions in the spectra have been properly fitted, leading to a non-optimal choice of parameters.} 

\begin{figure}
\begin{center}
\begin{tabular}{c}
\includegraphics[width=5.8cm,angle=270]{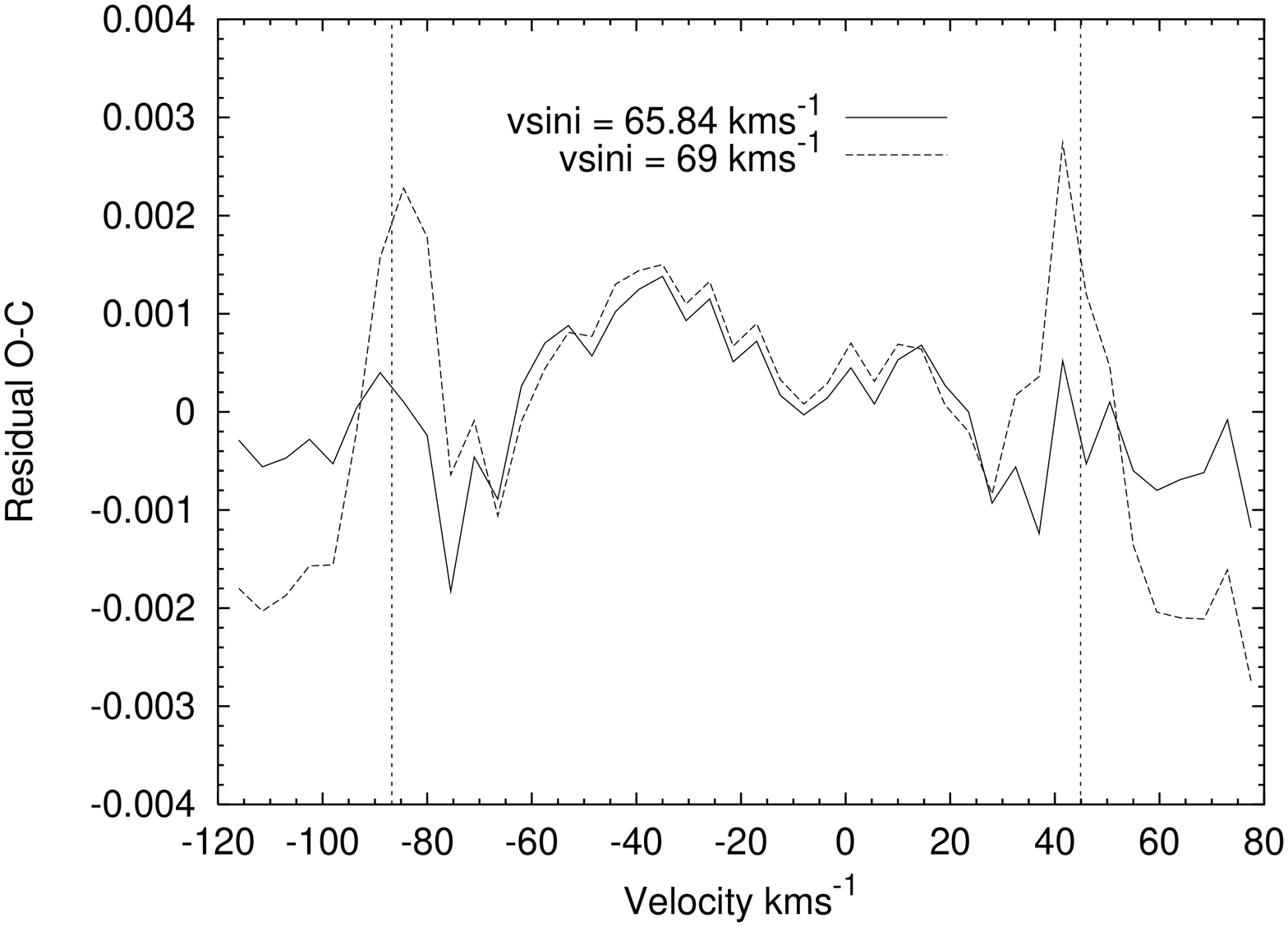} \\
\end{tabular}
\end{center}
\caption[Residual spots]{The mean residual plots from fits to the 1998 July 04 -10 data using \vsini\ = 65.84 \kms and \vsini\ = 69 \kms. The dashed lines indicate \vsini\ = 65.84 \kms. }
\protect\label{fig:residuals}
\end{figure}

{ We therefore determine the system parameters by minimising the value of \chisq\ attained after a fixed number of maximum entropy regularised iterations. The results are presented in Table \ref{tab:sysparam} (we do not tabulate the equivalent width which depends on the strength and number of lines used in the least squares deconvolution). Since we do not reach a reduced \hbox{\chisq\ = 1}, we use the TEST statistic \citep{skilling84} to determine the value of \chisq\ at which we begin fitting noise. Reasons for not attaining \hbox{\chisq\ = 1} can be attributed to incorrectly determined error bars (i.e. systematically too large or too small), and also to non-flat continuum on either side of the deconvolved rotation profiles. While the latter has negligible effect on the reconstructed data, incorrectly determined error bars have no effect other than to alter the optimal \chisq\ which can be achieved. 1-$\sigma$ uncertainties for each parameter are determined for \mbox{$\Delta\chi^2 = 1 + d/\chi^2_{min}$}, where $d$ is the number of fitted data points and \chisq$_{min}$ the lowest achieved \chisq. This assumes that mis-determination of noise is the only reason that \hbox{\chisq\ = 1} is not achieved  and will thus yield conservative estimates for parameter uncertainties.}

In \citet{barnes01aper} we also re-determined system parameters from the original study \citep{barnes98aper} of the \hbox{G dwarf} \hbox{He 699} using the same method as in the present study. There, we found that using spot area as a proxy for the image entropy tended to overestimate the \vsini\ and equivalent width, a finding shown in \citet{barnes99thesis}. In the case of He 699, a \hbox{\vsini\ = 96 \kms}\ was found using the spot area criterion, compared with \hbox{\vsini\ = 93.5 \kms}\ using the \chisq\ criterion. At \hbox{65.84 \kms}, the present result for LO Peg is 4.6\% lower, having important effects on the reconstructed image. Overestimation of \vsini\ means that the radial acceleration of starspot features cannot be correctly determined and has resulted in starspots being placed in a low latitude band.

Table \ref{tab:sysparam} shows the optimal axial inclination to be 45\degs. This is slightly lower than the 50\degs\ estimated by \citet{jeffries94bd22d4409} (and used by \citet{lister99lopeg}) but well within their 10\degs\ uncertainty. Attempts to reconstruct images at higher inclinations fail to fit the flat-bottomed profile correctly, because the polar region is too foreshortened to yield the necessary flux deficit, even when the spot occupancy saturates at 100\%. Fig. \ref{fig:residuals} is a plot of the mean residual line profile for maximum entropy fits using \vsini\ = 69 \kms\ and \hbox{65.84 \kms}\ indicating the lower residuals in the wings using the latter value. These wing residuals are also smaller for \vsini\ =  \hbox{65.84 \kms}\ when a similar plot is made for the 1993 data set. { The goodness of fit obtained for the 1998 data set with \vsini\ = 69 \kms\ and \vsini\ = 65.84 \kms\ is \hbox{\chisq$_{min}$ = 2.82} and \hbox{\chisq$_{min}$ = 2.34} respectively. If we assume that \hbox{\vsini\ = 65.84 \kms}\ is the optimally determined projected rotation velocity, we note that \vsini\ = 69 \kms\ falls outwith the 99.99\% confidence interval.}
 
 We have therefore recalculated an image of the 1993 data set using both nights of data (Fig. \ref{fig:lopeg_images2}) along with the combined image for the 1998 data in order to allow direct comparison of consistently determined images. The images from both epochs are now very similar in terms of the polar cap strength and location of starspots at lower latitudes, as indicated by the mean spot filling plot on the right of each map.  There is marginally less mean filling in the polar regions from 1993, but the low latitude features in both cases only occur at or above $\sim$25\degs. { The degree of similarity of the images at both epochs is most likely coincidental as the chosen ephemeris is different for each image reconstruction. Also, the error in the determined period results in an uncertainty of the time of zero phase by up to $\pm 5$ hrs in the five year interval between observations.}

\begin{figure}
\begin{center}
\begin{tabular}{c}
\includegraphics[height=8.5cm,width=8cm,angle=270,bbllx=25,bblly=0,bburx=530,bbury=745]{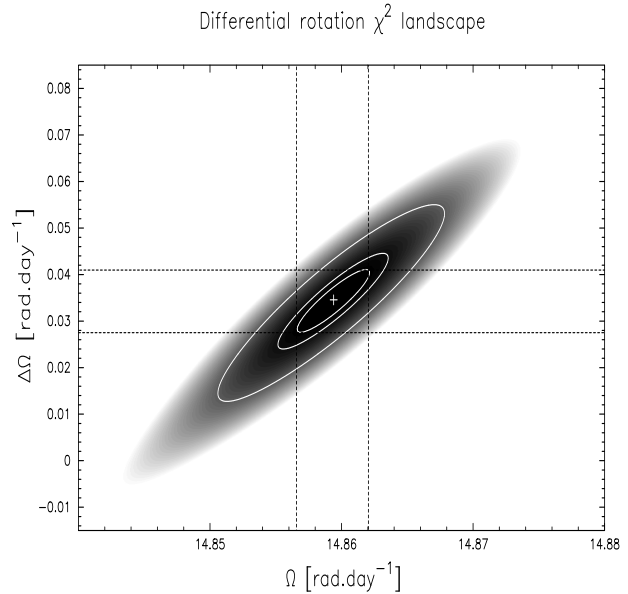} \\
\end{tabular}
\end{center}
\caption[Chisq landscape]{\chisq\ landscape for July 04 - 10 fitted data set. Shown are the \chisq\ minimum, marked by a cross, and from centre outwards, the 1-parameter 68.3\%, 2-parameter 68.3\% and 99.99\% confidence intervals. The dashed lines mark the 1-parameter 1-$\sigma$ or 68.3\% values for each paremeter. }
\protect\label{fig:diffrot}
\end{figure}

\subsection{Latitude dependent rotation}

The seven night timebase of the observations enable us to investigate the evolution of surface features on LO Peg.
Any given phase is observed approximately three times on alternate nights which should be sufficient to track starspot features over the timespan. Additionally, the images in \hbox{Fig. \ref{fig:lopeg_images}} indicate that most spots are sufficiently stable during the seven nights of observations to warrant such an investigation. We have discussed the inclusion of differential rotation in our Doppler imaging code in previous publications, a method first applied successfully by \citet{donati00rxj1508} to the \hbox{early G pre-main} sequence star RX J1508.6-4423. A solar type differential rotation law involving a shift term applied to the angular velocity \mbox{$\Omega_0$}, and a latitudinal dependent shear term, \mbox{$\Delta\Omega$}, are fitted according to \mbox{$\Omega(\theta) = \Omega_0 - \Delta\Omega\rm{sin}^2(\theta)$}. We minimise the fit in the same way as for other system parameters, by obtaining the lowest \chisq\ after a fixed number of iterations.

For LO Peg, we find that \hbox{\mbox{$\Omega_0 = 14.86$}} \hbox{\mbox{$\pm 0.0027 $} \radday}\ and \hbox{\mbox{$\Delta\Omega = 0.034714$}} \hbox{\mbox{$\pm 0.006692$} \radday.} With an axial rotation period, \hbox{P = 0.423229 d}, this implies an equivalent phase shear amplitude \hbox{\mbox{$\Delta\phi = 0.005525 \pm 0.001065$} \invday}, or an equator-pole lap time of \hbox{\mbox{$181 \pm 35$} d.}

\section{Discussion}

\subsection{Starspot distributions}
With a spectral type of K5 - K7, LO Peg is one of the latest spectral type dwarfs which have been Doppler imaged. Additionally, it is the only main sequence star in the \hbox{mid-late K} spectral type range to have been studied in this way. The evolution of starspots on the one-week timescale of the current observations is clearly quite small. Over the longer five year time base, the starspot morphology has also changed very little in terms of latitudinal distribution. The polar cap appears to have remained both stable and strong at both epochs of observation. This is intriguing because the slightly earlier spectral type (K3V), but more rapidly rotating \hbox{(P = 0.38 d)} Speedy Mic showed no evidence for a polar cap \citep{barnes01speedy} when observed at a single epoch in 1998. The early M dwarf stars, EK Dra (RE1816+541) and HK Aqr have similar periods to LO Peg but do not possess polar caps, at more than one epoch in the latter case \citep*{barnes04hkaqr}. There is however evidence that the young K0V star AB Dor \citep*{kurster94abdor} did not possess a polar cap in 1989, in which case these large spots may disappear periodically. If such behaviour is (quasi-)cyclic, it may be related to dynamo activity, although there is as yet no convincing evidence for this. 

By simulating a star ten times more active than the Sun, \citet{schrijver01polar} found that the higher rate of flux emergence, at around 30 times solar, resulted in the development of polar caps. But magnetic flux, as traced by starspots on LO Peg, clearly does not emerge at low latitudes in the way we see on the Sun. With the flux emergence patterns seen on young rapid rotators in mind, \citet{mackay04} have investigated the role of meridional flows for stars where flux emerges up to latitude 70\degs. In order to reproduce the intermingled polarity pattern as seen on the young rapid rotator, AB Dor, strong meridional flows of around 100 \ms\ are required. 

Unfortunately we do not have magnetic maps of LO Peg which would allow us to compare our results more closely, but the the flux distribution patterns do demand comparison with the simulation maps. \citet{mackay04} have shown that much of the flux piles up at high latitudes, with less at intermediate latitudes. In the case where the rate of flux emergence shows cyclic behaviour (with an 11 year cycle), but no change in the latitude at which flux erupts over that cycle, fewer spots are seen at low latitudes than intermediate and high latitudes. While the amount of flux at low latitudes varies throughout the cycle, features within a few degrees of the equator are present at all times. However when a solar-like variation of flux emergence with latitude, varying between $\pm$70\degs (rather than the solar $\pm$40\degs) to $\pm$ 5\degs, is simulated, there are times when no flux is present in the equatorial regions.  A distribution resembling this situation is certainly seen on the present images of LO Peg where we see no spots below 15\degs. If observation can be reconciled with this picture of activity, long term Doppler imaging projects are required. It is worth noting however that while \citet{baliunas95} found period variability, indicating the presence of  active regions at different latitudes and differential rotation, the younger stars in the Mount Wilson Survey tended to show no preferred period for magnetic cycles.

A solar type interface dynamo predicts only spots at intermediate latitude in rapid rotators (\citealt{schussler96buoy} and \citealt{granzer00}). { While spots {\em do} appear at intermediate latitudes in both 1993 and 1998, there is some structure at lower latitudes, while the images are dominated by polar or circumpolar spots. In many cases however, in addition to polar and circumpolar structure, spots my be seen in bands at low latitudes.} This is the case for the young post-T Tauri star, RX J1508.6-4423 \citep{donati00rxj1508} and Alpha Per G dwarfs \citep{barnes01aper} for example. The role of meridional flows have been touched upon above and are known, in the solar case at least, to possess maximum amplitude at equatorial latitudes (e.g. \citealt*{komm93meridional}). Flow rates following the solar pattern, but with greater amplitude, as modelled by \citet{mackay04}, will then result in a lower chance of seeing starspots at low latitudes. Another likely reason for no spots being observed at low latitudes is related to the low axial inclination of LO Peg. At 45\degs, spots in the equatorial latitudes of the star will appear severely foreshortened. If the S/N of the data is not sufficiently high, and only small spots exist, our imaging program may not find sufficient evidence for their presence.

\subsection{Differential rotation}
The differential rotation rate which we have measured for LO Peg is interesting in the context of other measurements. The value of equator-pole lap time, \hbox{$\delta$ =  \mbox{$181 \pm 35$} d}, lies on a trend (albeit sparsely populated) with spectral type. \citet{donati00rxj1508} found \hbox{$\delta$ = 50 $\pm$ 10 d} for the young G2 pre-main-sequence star RXJ 1508.6-4423, while \citet{marsden04} found \hbox{$\delta$ = 66 $\pm$} 14 d for the IC2602 G2 star, R58. Over a period of 8 years, \citet{cameron02twist} found that the K0 dwarf AB Dor exhibits systematic changes in differential rotation rate from one season to the next, with \hbox{$\delta$ =  70 d to 140 d}. At approximately the same spectral type, the near-one-day period PZ Tel \citep{barnes00pztel} exhibited an equator-pole lap time of 80 $\pm$ 15 d. The latest spectral type star for which we have a differential rotation measurement \citep{barnes04hkaqr} is the dM1-2 object, HK Aqr which exhibited an equator-pole lap time { formally consistent with zero} (i.e. between \hbox{-1449 d and +448 d.}) Although the sense of the differential rotation was not clear, the magnitude is essentially very small.

LO Peg clearly fits into this trend of generally decreasing differential rotation as we descend the main sequence from early G through to early M. The changing AB Dor differential rotation however shows that there can be significant change, possibly dependent on the phase of a magnetic cycle. In all cases, the magnitude of differential rotation appears, to first order at least, to be independent of rotation rate. The Sun for instance has an axial rotation period of \hbox{27 d} and exhibits an equator-pole lap time of 120 d, whereas this value falls only by a factor of two (twice the rate of differential rotation) for rapidly rotating objects with 1/50th the solar period. These two predictions of weak rotation dependence and a decrease with rotation rate are predicted by the fluid circulation models of \citet{kitchatinov99drot}, although the exact magnitude of the predicted differential rotation does not match the observation. Additionally, for a \hbox{1 d} axial rotation period K5V star, these models predict a poleward meridional flow rate of $\sim$ 10 \ms, an order of magnitude lower than those required by \citet{mackay04} in their simulations. Clearly it would be desirable to determine an estimate of the magnitude and direction of flows observationally.

\section{ACKNOWLEDGEMENTS}

The data in this paper were reduced using {\sc starlink} software packages.

\end{document}